\documentstyle[12pt]{article}

\topmargin -\headheight

\begin{document}

\title{Finite Size XXZ Spin Chain\\ 
with Anisotropy Parameter $\Delta = \frac{1}{2}$}
\author{ V.~Fridkin$^{\rm a}$,  Yu.~Stroganov$^{\rm b}$ and
 D.~Zagier$^{\rm c}$\\\\
{$^{\rm a}$ Research Institute for Mathematical Sciences}\\
{ Kyoto University, Kyoto 606, Japan}\\
{$^{\rm b}$ Institute for High Energy Physics}\\
{ Protvino, Moscow region, Russia}\\
{$^{\rm c}$ Max-Planck-Institut fur Mathematik}\\
{ Gottfried-Claren-Strasse 26, D-53225, Bonn, Germany}}

\maketitle

\begin{abstract}
We find an  analytic solution of the Bethe Ansatz equations (BAE) for the
special case of a finite XXZ spin chain with free boundary conditions
and with a complex surface field which provides for $U_q(sl(2))$ symmetry
of the Hamiltonian. 
More precisely, we find one nontrivial solution, corresponding to 
the ground state of the system with anisotropy parameter $\Delta =
\frac{1}{2}$ corresponding to $q^3 = -1$.

With a view to establishing an exact representation of the ground
 state of the finite size XXZ spin chain in terms of elementary functions,
 we concentrate on the crossing-parameter $\eta$ dependence around
 $\eta=\pi/3$ for which there is a known solution. 
 The approach taken involves the use of a physical solution $Q$ 
of Baxter's t-Q equation, corresponding to the ground state,
 as well as a non-physical solution $P$ of the same equation.
 The calculation of $P$ and then of the ground state derivative is covered.
  Possible applications of this derivative to the theory of percolation
 have yet to be investigated.   

As far as the finite XXZ spin chain
with periodic boundary conditions is concerned,
we  find a similar solution for an assymetric case which corresponds to
the 6-vertex model with a special  magnetic field.
For  this case  we  find the analytic value of the ``magnetic moment''
of the system in the corresponding state.  

\end{abstract}

\hfill \it Dedicated to Rodney Baxter \\
\hspace*{\fill}  on the occasion of his 60th birthday.
		
\vspace{0.5cm}

\section*{I. Introduction}
		
\rm It is widely accepted that the Bethe Ansatz equations for an  integrable
quantum spin chain can be solved analytically only in the thermodynamic
limit or for a small number of spin waves or short chains.
In this paper, however, we have managed to find a special solution of
 the BAE for a  spin chain of arbitrary length N with  N/2 spin waves.

It is well known (see, for example~\cite{ABBBQ} and references therein)
that there is a correspondence between the Q-state Potts Models and
the Ice-Type Models with anisotropy parameter $\Delta = \frac{\sqrt Q}{2}$.
The coincidence in the spectrum of an N-site self-dual Q-state quantum 
Potts chain with free ends with a part of the spectrum of the 
$U_q(sl(2))$ symmetrical 2N-site XXZ Hamiltonian (\ref{eq:b})
is to some extent a manifestation of this correspondence.
\begin{equation}
   \label{eq:b}
   H_{xxz}=\sum_{n=1}^{N-1}\{\sigma_n^{+}\sigma_{n+1}^{-}
   +\sigma_n^{-}\sigma_{n+1}^{+}
   +\frac{q+q^{-1}}{4}\,\sigma_n^z\sigma_{n+1}^z\
   +\frac{q-q^{-1}}{4}\,(\sigma_n^z-\sigma_{n+1}^z)\},
\end{equation}
where $\Delta=(q+q^{-1})/2$.
This Hamiltonian was considered by  Alcaraz {\it et al.}~\cite{ABBBQ}
and its $U_q(sl(2))$ symmetry was described  by Pasqier and 
Saleur~\cite{PS}.
The family of commuting transfer-matrices that commute with 
 $H_{xxz}$  was constructed by 
Sklyanin~\cite{SKL} incorporating a  method of Cherednik~\cite{CHER}. 

Baxter's T-Q equation for the case under consideration can be written 
as~\cite{ZH}  
\begin{equation}
 \label{eq:g}
 t(u)Q(u)=\phi(u+\eta/2)Q(u-\eta) + \phi(u-\eta/2)Q(u+\eta)
\end{equation}  
where $q=\exp i\eta,\> \phi(u)=\sin 2u\,\sin^{2N}u$ and $t(u)=\sin 2u\,T(u)$.
The $Q(u)$  are  eigenvalues of Baxter's auxilary matrix $\hat Q(u)$,
where $\hat Q(u)$ commutes with the transfer matrix $\hat T(u)$.
The eigenvalue $Q(u)$ corresponding to an eigenvector with $M=N/2-S_z$
 reversed spins has the form
\begin{equation}
 \label{eq:h}
 Q(u)= \prod\limits_{m=1}^M \sin (u-u_m) \sin (u+u_m).
\end{equation}  
Equation (\ref{eq:g}) is equivalent to the Bethe Ansatz equations~\cite{MN}
\begin{equation}
\label{eq:i}
\left[\frac{\sin (u_k+\eta/2)}{\sin (u_k-\eta/2)}\right]^{2N}=
\prod\limits_{m\not=k}^M \frac{\sin (u_k-u_m+\eta)\sin (u_k+u_m+\eta)}
   {\sin (u_k-u_m-\eta)\sin (u_k+u_m-\eta)}.
\end{equation}  

Baxter's equation  can be interpreted  as a discrete version of
 a second order differential equation\cite{BLZ,PRST}.
So we can look for its second independent  solution $P(u)$ with
the same eigenvalue $T(u)$. 

In a recent article~\cite{BS} Belavin and Stroganov argued 
that the criteria for the above 
mentioned correspondence is the existence of a  second trigonometric 
solution for Baxter's T-Q equation and it was shown that in the case
$\eta=\pi /4$ the spectrum of $H_{xxz}$  contains the spectrum 
of the Ising model.
In this article we limit ourselves to the case $\eta = \pi /3$.
This case is in some sense  trivial since for this value of $\eta$, 
 $H_{xxz}$  corresponds to the 1-state Potts Model. 
We find  only one eigenvalue $T_0(u)$ of the transfer-matrices 
$\hat T(u)$ when Baxter's equation (\ref{eq:g}) has two independent
trigonometric solutions.
Solving for $ T(u)=T_0(u)$ analytically we find 
a trigonometric polynomial $Q_0(u)$ the zeros of which satisfy the
 Bethe Ansatz
equations (\ref{eq:h}). The number of spin waves is equal to $M=N/2$.
The corresponding eigenstate is the groundstate of $H_{xxz}$ with
 eigenvalue $E_0 = \frac{3}{2}(1-N)$, as discovered by Alcaraz 
{\it et al.}~\cite{ABBBQ} numerically.   

 When the first version of this paper
~\cite{FSZ} was reported at the Workshop in ANU (February 2000), Baxter
showed us his article~\cite{Baxter} where he writes about the existence of a
simple eigenvalue of the T-matrix for the 8-vertex model for the  special case
$\mu = \pi /3$.
About 30 years ago Baxter discovered~\cite{BXYZ} that the ground state energy
of the XYZ-spin chain which is described by the Hamiltonian
\begin{equation}
   \label{eq:xyz}
   H_{xyz}=-\frac{1}{2}\sum_{n=1}^{N}\{J_x \sigma_n^{x}\sigma_{n+1}^{x}
   +J_y \sigma_n^{y}\sigma_{n+1}^{y}
   +J_z \sigma_n^z\sigma_{n+1}^z\},
\end{equation}
has an especially simple value $E=-(J_x+J_y+J_z)/4$ for the case 
$\mu =\pi /3$, when $J_x,J_y$ and $J_z$ satisfy the
 constraint $J_x J_y + J_y J_z + J_z J_x =0$.        
This value was found in the thermodynamic limit only.
Later  in 1989~\cite{Baxter}
Baxter considered  a determinant functional relation and found a very simple
solution  for eigenvalue of T-matrix (see Section II below) for the case 
$\mu = \pi /3$.

There are however problems with the realization of this 
simple solution.
If we consider, for example, the usual XXZ-spin chain with periodic boundary
conditions we will find, solving  Baxter's T-Q equation, that 
the degree of Q(u) is $N/2+1$
(N even) and there is no Bethe vector for this simple solution.
We believe that for the case without a field, a ``physical''
solution  Q(u) must have degree less than or equal to $N/2$.
Baxter informed us that the addition of a special field saves
 the simple solution and
below we find the corresponding Q(u).

At  first glance it seems we have a clear picture.
For the open chain with special boundary conditions we have  central charge
$c=0$~(see for example~\cite{Mur}), so finite-size corrections are absent
and we have a simple solution for the ground state.
On the other hand, for periodic boundary conditions without a field
we have due to Hamer~\cite{Ham} $c \ne 0$, so there is  no simple solution.

We wish to stress that essentially only chains with even length were
 considered.
Hamer~\cite{Ham} considered only even N.  However, for odd N,
 one can find~\cite{YGS} a simple eigenvalue $T_0=(a+b)^N$ for the case when
the weights of the 8-vertex model satisfy  the condition
\begin{equation}
(a^2+a b) (b^2+a b) = (c^2 + a b) (d^2 + a b),
\end{equation}

which is equivalent  the condition $\mu = \pi /3$.

\section*{II. Consequences of the existence of the second ``independent''
 periodic solution}	

This question  was considered in article~\cite{BS}. Here we use a variation
more convenient for our goal.
This section is mainly due to Baxter~\cite{Baxter} (see also~\cite{Pearce}).

Let us consider T-Q equation (\ref{eq:g}) for $\eta = \frac{\pi}{L}$, where
$L \ge 3$ is an integer.
Let us fix a sequence of spectral parameter values $v_k = v_0 + \eta k$,
where k are integers and write
 $\phi_k=\phi(v_k-\eta/2)$, $Q_k=Q(v_k)$ and $t_k=t(v_k)$.
Functions $\phi(u)$, $Q(u)$ and $t(u)$ are periodic with  period  $\pi$.
 So the sequences we have introduced are also periodic with  
period $L$, i.e.,  $\phi_{k+L}=\phi_k$, etc..

Setting $u=v_k$ in (\ref{eq:g}) gives the linear system
\begin{equation}
\label{eq:gd}
 t_k Q_k=\phi_{k+1} Q_{k-1} + \phi_{k} Q_{k+1}.
\end{equation} 
 The matrix of coefficients for this system has the three-diagonal form
\begin{equation}
\label{eq:mat}
M=\begin{array}{|cccccc|}
-t_0   & \phi_0 &    0   & \dots & 0 & \phi_1 \\
\phi_2 &  -t_1  & \phi_1 & \dots & 0 &  0     \\
  0    & \phi_3 &  -t_2  & \dots & 0 &  0     \\
  .    &    .   &    .   & \dots & . &  .     \\
  .    &    .   &    .   & \dots & . &  .     \\
\phi_{L-1} & 0  &    0   & \dots &\phi_0 & -t_{L-1}     \\
\end{array}\>.
\end{equation}
Taking $v_0 \ne \frac{\pi m}{2}$, where $m$ is an integer, we have 
$\phi_k \ne 0$ for all $k$.

It is straightforward to calculate the determinant of the 
$L-2 \times L-2$ minor obtained  by  deleting the two left most columns
 and two lower most rows. It is equal to the product 
$-\phi_1^2 \phi_2 \phi_3 \dots \phi_{L-1}$, which is nonzero,
hence  the rank of  $M$ cannot be less than $L-2$.
Here we are interested in the case when the rank of M is precisely $L-2$ and
 we have two linearly independent solutions for  equation (\ref{eq:gd}). 
Let us  consider  the three simplest cases L = 3, 4 and 5.
Parameter $\eta$ is equal to $\frac{\pi}{3}$, $\frac{\pi}{4}$
and $\frac{\pi}{5}$ respectively.

For $L=3$ the rank of $M$ is unity  and we immediately get
$t_0=-\phi_2,\> t_1=-\phi_0$ and $ t_2=-\phi_1$.
Returning to the functional form, we can write
\begin{equation}
\label{eq:t}
T_0(u)=t_0(u)/\sin 2u=-\phi (u+\frac{\pi}{2})/\sin 2u = \cos^{2N} u.
\end{equation} 
 This is the unique eigenvalue of the transfer-matrix for which the 
T-Q equation has two ``independent'' periodic solutions.
  It is well known (see, for example,~\cite{MN}) that the eigenvalues of 
$H_{xxz}$ are related to the eigenvalues $T(u)$ by
\begin{equation}
\label{eq:en}
E=-N \cos \eta + ( \frac{T^{\prime}(\eta /2)}{T(\eta /2)}+\tan \eta) \sin \eta.
\end{equation} 
For the eigenstate corresponding to eigenvalue (\ref{eq:t}) we obtain
\begin{equation}
\label{eq:en0}
E_0=\frac{3}{2}(1-N).
\end{equation} 
This is the groundstate energy which was
 discovered  by Alcaraz {\it et al.}~\cite{ABBBQ} numerically.   

In the next section we find all solutions of Baxter`s T-Q equation 
corresponding to $T(u)=T_0(u)$. 
In particular we find  $Q(u)$, the  zeros of which
 satisfy the BAE (\ref{eq:i}).

For  $\eta = \frac{\pi}{3}$ and transfer-matrix eigenvalue 
 $T_0(u)=\cos^{2N} u$, T-Q equation (\ref{eq:g}) reduces to 
\begin{equation}
\label{eq:phiQ}
\phi(u+3\eta/2) Q(u) + \phi(u-\eta/2) Q(u+\eta) + \phi(u+\eta/2) Q(u-\eta)=0.
\end{equation} 

Keeping $t(u)$ arbitrary, for the  moment, we can rewrite (\ref{eq:g}) as
\begin{equation}
t(u)=\phi(u+\eta /2) \frac{Q(u-\eta)}{Q(u)} + 
\phi(u-\eta /2) \frac{Q(u+\eta)}{Q(u)}.
\nonumber
\end{equation}
We also have
\begin{equation}
t(u+\eta)=\phi(u+3\eta /2) \frac{Q(u)}{Q(u+\eta)} + 
\phi(u+\eta /2) \frac{Q(u+2\eta)}{Q(u+\eta)}.
\nonumber
\end{equation}
Multiplying these equations together we obtain the fusion relation
\begin{equation}
\label{eq:fr}
t(u) t(u+\eta) = \phi(u-\eta /2) \phi(u+3\eta /2) + \phi(u+\eta/2) \tilde t(u),
\end{equation}
where
\begin{equation}
\tilde t(u) = \frac{Q(u-\eta)}{Q(u) Q(u+\eta} 
(\phi(u+3\eta /2) Q(u) + \phi(u-\eta /2) Q(u+\eta) 
+ \phi(u+\eta /2) Q(u-\eta)).
\nonumber
\end{equation}
In the case under consideration we have $\tilde t(u) = 0$.
Fusion relation (\ref{eq:fr}) reduces to the simple equality
\begin{equation}
\label{eq:ir}
t_0 (u) t_0 (u+\eta) = \phi(u-\eta /2) \phi(u+3\eta /2),
\end{equation}
which is a kind of inversion  relation.

 For $L=4$
\begin{equation}
\label{eq:mat4}
M=\begin{array}{|cccc|}
-t_0   & \phi_0 &    0   &  \phi_1 \\
\phi_2 &  -t_1  & \phi_1 &   0     \\
  0    & \phi_3 &  -t_2  &  \phi_2     \\
\phi_3 & 0      &\phi_0  & -t_3     \\
\end{array}\>.
\end{equation}
Deleting the second row and the forth column we obtain a minor with 
determinant $-\phi_0 \phi_3 (t_0+t_2)$. It is zero when $t_2=-t_0$, i.e.,
 $t(u+\frac{\pi}{2})=-t(u)$.
Considering the other minors we obtain the functional equation
\begin{equation}
\label{eq:E}
t(u+\pi/8)t(u-\pi/8)=\phi(u+\pi/4)\phi(u-\pi/4)-
\phi(u)\phi(u+\pi/2).   
\end{equation}
This functional equation was used in~\cite{BS} to find  $t(u)$
and  show that this part of the spectrum of $H_{xxz}$ coincides with the
 Ising model.
It would be interesting to find a corresponding $Q(u)$.

Lastly for $L=5$, minor $M_{35}$ (the third row and the fifth column are 
deleted) has determinant 
$\phi_0 \phi_4 (t_0 t_1 + \phi_1 t_3 - \phi_0 \phi_2)$.
Setting this to zero we have
\begin{equation}
\label{eq:E5}
t(u) t(u+\pi/5) + \phi(u+\pi/10) t(u+3\pi/5)
 - \phi(u-\pi/10) \phi(u+3 \pi/10)=0.
\end{equation}
It is not difficult to check that in this case all  $4 \times 4$ minors have
zero determinant and that the rank of M is 3. Thus we have two ``independent''
periodic solutions of Baxter's T-Q equation.

Note that this functional relation (\ref{eq:E5}) coincides with the 
Baxter-Pearce relation for the  hard hexagon model.
The connection between (\ref{eq:E5}) and a special value of the rank
of the matrix of coefficients for system (\ref{eq:gd}) was remarked upon 
in~\cite{ABF} by Andrews, Baxter and Forrester.

For general $L$ we  obtain the same truncated functional relations that
have been obtained in \cite{BS} with the same assumptions.
Note that for the ABF models~\cite{ABF}, which are a generalization 
of the hard hexagon model, the truncated functional relations have been
proved by  Behrend, Pearce and O'Brien in~\cite{BPO}.

\section*{III. Solution of Baxter's Equation for $\eta = \frac {\pi}{3}$
and $T=T_0$}	 

Equation (\ref{eq:phiQ}) can be rewritten as follows:
\begin{equation}
\label{eq:no3}
f(v)+f(v+\frac{2\pi}{3})+f(v+\frac{4\pi}{3}) = 0,
\end{equation}
where $f(v) = \sin v\> \cos^{2N} (v/2)\> Q(v/2)$ has period $2\pi$.
The trigonometric polynomial $f(v)$ is an odd function so it can be written
\begin{equation}
\label{eq:ff}
f(v) = \sum_{k=1}^{K} c_k \sin kv,
\end{equation}
where $K$ is the degree of $f(v)$.
Equation (\ref{eq:no3}) is equivalent to
\begin{equation}
\label{eq:no3p}
c_{3m}=0, \qquad m \in  Z.
\end{equation}
The point $v=\pi$ is a zero of $f(v)$  of order $2N+1$, so we obtain
\begin{equation}
(\frac{d}{dv})^i f(v)|_{v=\pi} = 0,\qquad i = 0,1,\dots ,2N.
\nonumber
\end{equation}
For even $i$ this condition is immediate, whereas for $i=2j-1$ we  use  
(\ref{eq:ff}) to obtain
\begin{equation}
\label{eq:ls}
\sum_{k=1, k \ne 3m}^{K} (-1)^k c_k k^{2j-1} = 0,\qquad j=1,2,\dots ,N.
\end{equation}

Our problem is a special case of a more general problem which can be formulated
as follows.
Given a set of different complex numbers $X=\{x_1,x_2,\dots,x_I\}$
we seek another complex set $B=\{\beta_1,\beta_2,\dots,\beta_I\}$
where $\beta_i \ne 0$ for some $i$, so that 
\begin{equation}
\label{eq:prob}
\sum_{i=1}^{I} \beta _i P(x_i) = 0
\end{equation}
for any polynomial $P(x)$ of degree not more than $N-1$. 
It is clear that for $I \le N$ the system $B$ does not exist.
If $\beta_1 \ne 0$, for example, the product 
$(x-x_2)(x-x_3)\dots(x-x_I)$ provides a counterexample.

Let $I=N+1$. We try the polynomials
\begin{equation}
\label{eq:pol}
P_{r} = \prod_{i=1, i \ne r,}^{N} (x-x_i), \qquad r=1,2,\dots, N.
\end{equation}  
Condition (\ref{eq:prob}) gives
$\beta_r P_{r}(x_r) + \beta_I P_{r}(x_I) = 0$
and we immediately obtain 
\begin{equation}
\label{eq:sol}
\beta_r = \mbox{const} \prod_{i=1, i \ne r}^{N+1}(x_r-x_i)^{-1},
\end{equation}
which is a solution  because  the system (\ref{eq:pol}) forms 
a basis of the linear space of $N-1$ degree polynomials. 
So for $I=N+1$ we have a unique solution
 (up to an arbitrary nonzero constant) given by (\ref{eq:sol}).
 
Returning to (\ref{eq:ls}) we first consider  $N=2n$,
$n$ a positive integer. Fix $I=N+1=2n+1$. The degree $K$ becomes $3n+1$. 
It is convenient to use a new index
$k=|3\kappa+1|$, where $|\kappa| \le n$. 
Equation (\ref{eq:ls}) can be rewritten as 
\begin{equation}
\label{eq:newls}
\sum_{\kappa=-n}^{n} \beta_{\kappa}(3\kappa+1)^{2(j-1)}=0,
 \qquad j=1,2,\dots,N,
\end{equation}
where we use new unknowns
$\beta_{\kappa}=(-1)^{\kappa} c_{|3\kappa +1|} |3\kappa+1|$
instead of $c_k$.
Using  (\ref{eq:sol}) we obtain 
\begin{equation}
\label{eq:sol2}
\beta_{\kappa} = \mbox{const} \prod_{\rho=-n, \rho \ne \kappa}^{n}
((3\kappa+1)^2-(3\rho+1)^2)^{-1}.
\end{equation}
We can rewrite this using  binomial coefficients as
\begin{equation}
\label{eq:sol3}
\beta_{\kappa} = \mbox{const}^{\prime}\>(\kappa+\frac{1}{3})
\biggl(\begin{array}{c}
2n + \frac{2}{3} \\
n - \kappa
\end{array}\biggr)\biggl(\begin{array}{c}
2n - \frac{2}{3} \\
n + \kappa
\end{array}\biggr).
\end{equation}
The old system of unknowns is given (up to an arbitrary constant) by
\begin{equation}
\label{eq:sol4}
c_{3\kappa+1} = (-1)^{\kappa}\mbox{ sgn}\>(\kappa + \frac{1}{3})
\biggl(\begin{array}{c}
2n + \frac{2}{3} \\
n -\kappa
\end{array}\biggr)\biggl(\begin{array}{c}
2n - \frac{2}{3} \\
n + \kappa
\end{array}\biggr)
\end{equation}
and using (\ref{eq:ff}) we obtain the function $f(v)$,
\begin{equation}
\label{eq:sol5}
f(v)  = \sum_{\kappa = -n}^{n}(-1)^{\kappa}
\biggl(\begin{array}{c}
2n + \frac{2}{3} \\
n - \kappa
\end{array}\biggr)\biggl(\begin{array}{c}
2n - \frac{2}{3} \\
n + \kappa
\end{array}\biggr) \sin (3\kappa + 1)v.
\end{equation}

We recall that the solution of Baxter's T-Q equation for $T(u)=T_0(u)$
 is given by
\begin{equation} 
Q_0 (u) = \frac{f(2u)}{\sin 2u \cos^{2N} u}
\end{equation}
and its zeros $\{u_k\}$ satisfy the BAE (\ref{eq:i}).  

\section*{IV. Distribution of zeros of $Q_0 (u)$}	 

Function $f(v)$ (\ref{eq:sol5}) can be shown to satisfy
a simple second order linear differential equation. The coefficient functions
of this  equation  are closely connected with the density function  of the 
zeros of $Q_0(u)$ in the thermodynamic limit.

Let us introduce  $x=\exp{(3iv)}$ and rewrite $f(v)$ as
\begin{equation} 
f(v) = F(x) = F^{+}(x)-F^{-}(x),\qquad F^{+}(x) = \sum_{\kappa=-n}^{n} 
a_{\kappa} x^{\kappa +  \frac{1}{3}}, \qquad F^{-}(x)=F^{+}(1/x).
\end{equation}
We have a multiplicative recursion relation for $a_{\kappa}$
\begin{equation} 
\frac{a_{\kappa+1}}{a_{\kappa}} = -\frac{(n-\kappa)(n-\kappa - 2/3)}
{(n+\kappa +1)(n+\kappa + 5/3)}
\end{equation}
which gives \footnote{$F^{+}(\kappa)=F(-2n,2/3-2n,5/3,-x) x^{1/3-n}$,
where $F(a,b,c,x)$
is the usual Gauss hypergeometric function.}
\begin{equation} 
\sum_{\kappa=-n}^{n}
\{a_{\kappa+1}(n+\kappa +1)(n+\kappa + 5/3)+
a_{\kappa}(n-\kappa)(n-\kappa - 2/3)\}x^{\kappa+\frac{1}{3}}=0
\end{equation}
or
\begin{equation} 
\sum_{\kappa=-n}^{n}
a_{\kappa}\{(n+\kappa + 1/3)^2 - 1/9\}x^{\kappa-\frac{2}{3}}+
\sum_{\kappa=-n}^{n}
a_{\kappa}\{(n-\kappa - 1/3)^2 - 1/9\}x^{\kappa+\frac{1}{3}}=0.
\end{equation}
Using the operator $\theta=x\frac{d}{dx}$ we can rewrite this as 
\begin{equation}
\label{eq:dif} 
\{((\theta + n)^2 - 1/9)/x + 
(\theta - n)^2 - 1/9
\} F^{+}=0.
\end{equation}
This equation is invariant under the transformation  $x \rightarrow 1/x$,
 so $F^{-}(x)=F^{+}(1/x)$ is also a solution for (\ref{eq:dif}).

Alternatively, we know that in a neighbourhood around  the singular point
 $x=-1$  there are two solutions of (\ref{eq:dif}) which behave as 
$(x+1)^\alpha (1+O(x+1)+\dots)$. Substituting this into 
(\ref{eq:dif}),
we obtain  $\alpha_1=0$ and $\alpha_2=4n+1$.
This method can be used as an another  way to find $f(v)$.

Returning to variable $v$ we have
\begin{equation}
\label{eq:dift} 
\{\exp (-3iv)((d/dv +3i n)^2 + 1) + 
((d/dv -3i n)^2 + 1)\} f(v)=0.
\end{equation}
Multiplication  with $\exp (3iv/2)/\cos (3v/2)$ gives
\begin{equation}
\label{eq:diftb} 
\frac{d^2f}{dv^2} + 6n\tan (3v/2)\frac{df}{dv} + (1 - 9 n^2)\>f=0.
\end{equation}

The zeros of $f(v)$, the density of which is important in the thermodynamic
limit, are located on the imaginary axis in the complex $v$-plane.
So it is convenient to make the change of variable $v=i s$.
It is also useful to introduce another function 
$g(s) = f(is)/\cosh^{2n}(3s/2)$.
The differential equation for $g(s)$ is then
\begin{equation}
\label{eq:diftbb} 
g^{\prime \prime} + \left(\frac{9n(2n+1)}{2\cosh^2(3s/2)}-1\right)g = 0.
\end{equation}
Let $g(s_0)=0$. For large $n$ we have in a small vicinity of $s_0$
\begin{equation}
\label{eq:difas} 
g^{\prime \prime} + \left(\frac{3n}{\cosh(3s_0/2)}\right)^2 g = 0.
\end{equation}
This equation describes a  harmonic oscillator with  frequency 
$\omega_0=3n/\cosh(3s_0/2)$.
The  distance between nearest zeros is approximately 
$\Delta s=\pi/\omega$ and we obtain the following density function
 which describes the  number of zeros per unit length 
\begin{equation}
\label{eq:den} 
\rho(s)=1/\Delta s=\omega/\pi=\frac{3n}{\pi \cosh(3s/2)}.
\end{equation}

It is possible to use the theory of Sturm-Liouville operators  for a more
careful consideration of (\ref{eq:diftbb}).
 We note that this equation has a history as rich as the 
BAE.  Eckart~\cite{Eck30} used the Schrodinger equation
 with a bell-shaped potential $V(r) = -G/\cosh^{2}r$ for phenomenological
 studies in atomic and molecular physics. Later it was used in chemistry,
biophysics and  astrophysics, just to name a few. For more recent references
 see for example~\cite{Zn}.  

\section*{V. $\eta$-dependence of the ground state 
energy of the finite XXZ spin chain}

We recall Baxter's t-Q equation in terms of the second solution $P$,
\begin{equation}
\label{eq:tP}
 t(u)P(u)=\phi(u+\eta/2)P(u-\eta) + \phi(u-\eta/2)P(u+\eta)
\end{equation}  

$P(u)$ is characterised by having degree $N+2$. 
  As in the case of $Q(u)$ we set up a function 
$f(v) = \sin v\> \cos^{2N} (v/2)\> P(v/2)$, and write 
\begin{equation}
\label{eq:fff}
f(v) = \sum_{k=1}^{K} c_k \sin kv,
\end{equation}
where $K$ is now $3n+2$.  The number of equations for coefficients
 $c_k$ is determined by $t(u)$ and is $N$ as before.
Thus in this case we have freedom in two of the $c_k$.


 Solving equation (\ref{eq:ls}) with this extra freedom,
 we obtain $f(v)=\alpha f_P(v) + \beta f_Q(v)$ where 
$\alpha$, $\beta$ are arbitrary, $f_Q(v)$ is the solution for $Q(u)$
 given by (\ref{eq:sol5}) and 
\begin{equation}
\label{eq:sol55}
f_P(v)  = \sum_{\kappa = -n}^{n}(-1)^{\kappa}
\biggl(\begin{array}{c}
2n + \frac{4}{3} \\
n - \kappa
\end{array}\biggr)\biggl(\begin{array}{c}
2n - \frac{4}{3} \\
n + \kappa
\end{array}\biggr) \sin (3\kappa + 2)v.
\end{equation}  
	 
Baxter's equation  can be interpreted  as a discrete version of
 a second order differential equation, so we can express its coefficients
in terms of the two independent solutions~\cite{BLZ,PRST}:
\begin{equation}
\label{eq:wron}
\phi(u)=P(u+\eta /2) Q(u-\eta /2) -P (u-\eta /2) Q(u+\eta /2)\> \equiv \>
p_{+} q_{-} - p_{-} q_{+}  
\end{equation}
and 
\begin{equation}
\label{eq:wrt}
t(u)=P(u+\eta) Q(u-\eta) -P (u-\eta) Q(u+\eta )\>  \equiv \>
P_{+} Q_{-} - P_{-} Q_{+}.  
\end{equation}

Consider now the functions $t,Q$ and $P$ as functions of two variables
$u$ and $\eta$.
$\phi(u)$ does not depend on $\eta$, so taking the derivative w.r.t. $\eta$
from (\ref{eq:wron}) we obtain
\begin{equation}
\label{eq:var1}
0=\delta p_{+} q_{-} + p_{+} \delta q_{-} - 
\delta p_{-} q_{+} - p_{-} \delta q_{+} + \frac{1}{2} 
(p_{+}^{\prime} q_{-} - p_{+} q_{-}^{\prime}
+p_{-}^{\prime} q_{+} - p_{-} q_{+}^{\prime}).  
\end{equation}
We use $\delta$ for the derivative w.r.t. $\eta$ and  $\prime$
for the derivative w.r.t. $u$.
Equation  (\ref{eq:wrt}) leads to
\begin{equation}
\label{eq:var2}
\delta t=\delta P_{+} Q_{-} + P_{+} \delta Q_{-} - 
\delta P_{-} Q_{+} - P_{-} \delta Q_{+} +  
P_{+}^{\prime} Q_{-} - P_{+} Q_{-}^{\prime}
+P_{-}^{\prime} Q_{+} - P_{-} Q_{+}^{\prime}  
\end{equation}
For $\eta = \pi /3$ we have, for example, $p_{+} \equiv P(u+\pi /6) =
P(u+\pi/ 2 - \pi /3)$,
so if we shift the variable $u$ in (\ref{eq:var1}) by $\pi /2$ we
get
 \begin{equation}
0=\delta P_{-} Q_{+} + P_{-} \delta Q_{+} - 
\delta P_{+} Q_{-} - P_{+} \delta Q_{-} + \frac{1}{2} 
(P_{-}^{\prime} Q_{+} - P_{+} Q_{-}^{\prime}
+P_{+}^{\prime} Q_{-} - P_{-} Q_{+}^{\prime})  
\end{equation}
Adding this equation to (\ref{eq:var2}) we arrive at 
the formula for the derivative of the largest eigenvalue of the transfer matrix
\begin{equation}
\label{eq:vart}
\delta t\> \equiv \>\left. \frac{\partial t}{\partial \eta}\right|_{\eta=\pi/3}
=\frac{3}{2}(P_{+}^{\prime} Q_{-} - P_{+} Q_{-}^{\prime}
+P_{-}^{\prime} Q_{+} - P_{-} Q_{+}^{\prime})  
\end{equation}
One can use  (\ref{eq:sol5}) and (\ref{eq:sol55}) to get an explicit
expression. 

Note that ``Wronskian'' (\ref{eq:wron}) reduces to a complicated combinatorial
identity
\begin{equation}
A(-K-1)-A(K-1) = (-1)^{K}\frac{K}{2n+1}\biggl(\begin{array}{c}
2n - \frac{2}{3} \\
2n
\end{array}\biggr)
\biggl(\begin{array}{c}
2n - \frac{4}{3} \\
2n
\end{array}\biggr)
 \biggl(\begin{array}{c}
4n + 2 \\
K+2n +1
\end{array}\biggr),
\end{equation}
where K is an integer $1-2n \le K \le 2n-1$
and 
\begin{equation}
A(K)\> \equiv \>\sum_{\kappa=-n}^{n}\sum_{\tilde \kappa =-n}^{n} 
\biggl(\begin{array}{c}
2n + \frac{2}{3} \\
n - \kappa
\end{array}\biggr)
\biggl(\begin{array}{c}
2n - \frac{2}{3} \\
n + \kappa
\end{array}\biggr)
 \biggl(\begin{array}{c}
2n + \frac{4}{3} \\
n -\tilde \kappa
\end{array}\biggr)\biggl(\begin{array}{c}
2n - \frac{4}{3} \\
n + \tilde \kappa
\end{array}\biggr)\>\delta_{\kappa+\tilde \kappa, K}. 
\end{equation}

\section*{VI. Finite XXZ spin chain with periodic boundary conditions
and with additional magnetic field}
The assymetric 6-vertex model under consideration was investigated by 
Baxter~\cite{Baxter6}. The associated spin chain Hamiltonian can be found in 
the paper 
of Perk and Schultz~\cite{Perk}.
Baxter's T-Q equation with additional field can be written as follows
\begin{equation}
\label{eq:TQmag}
T(u)\>Q(u) = \exp(h) \sin^N(u-\eta/2)\> Q(u+\eta)+\exp(-h) 
\sin^N(u+\eta/2)\> Q(u-\eta).
\end{equation}
Using a similar procedure to Section II (see also Baxter's original 
paper~\cite{Baxter})
we find that for $\eta = 2\pi/3$
or for $\eta = \pi/3$ we again have a simple solution for a special choice
of the field.
Let us fix $\eta = 2\pi/3$ and $h = i \eta$. 
For this case the transfer matrix T has an eigenvalue $-\sin^n u$.

Using the method of Section III, we find that 
\begin{eqnarray}
&f(u)\>\equiv\>\sin^N u\>Q(u) =      
\displaystyle\sum_{\kappa=0}^{n} 
\biggl(\begin{array}{c}
n - \frac{1}{3} \\
\kappa
\end{array}\biggr)
\biggl(\begin{array}{c}
n - \frac{2}{3} \\
n - \kappa
\end{array}\biggr)
\exp (i u (6\kappa - 3n))& - \nonumber \\
&-\displaystyle\sum_{\kappa=0}^{n-1} 
\biggl(\begin{array}{c}
n - \frac{1}{3} \\
\kappa
\end{array}\biggr)
\biggl(\begin{array}{c}
n - \frac{2}{3} \\
n - 1- \kappa
\end{array}\biggr)
\exp (i u (-6\kappa + 3n -2))&
\end{eqnarray}
where $n=N/2$.

The trigonometric polynomial $P(u) \equiv Q(-u)$ satisfies a conjugate 
equation which can be obtained from (\ref{eq:TQmag}) by changing the sign of 
 parameter $h$. We easily obtain (see for example~\cite{BLZ})
the analogues of (\ref{eq:wron}) and (\ref{eq:wrt}):
\begin{equation}
\label{eq:wronmag}
\phi(u)=\exp(-h)P(u+\eta /2) Q(u-\eta /2) -
\exp(h)P(u-\eta /2) Q(u+\eta /2),  
\end{equation}
and 
\begin{equation}
\label{eq:wrtmag}
T(u)=\exp(-2h)P(u+\eta) Q(u-\eta) - \exp(2h)P(u-\eta) Q(u+\eta ).
\end{equation}

Using similar procedure to the previous section we find the
derivative of eigenvalue $T(u)$ w.r.t. parameter $h$, which is the 
``magnetic moment'' we refered to in the abstract. 
\begin{equation}
\label{eq:dermag}
\left. \frac{\partial T}{\partial h}\right|_{h=2i\pi/3}=
-3\{\exp(h)\>Q(-u-\eta) Q(u-\eta) + \exp(-h)\>Q(-u+\eta) Q(u+\eta)\}.
\end{equation}

\section*{Acknowledgements}
We are grateful to  M. T. Batchelor, R. J. Baxter, V. V. Bazhanov,
 A. A. Belavin, L. D. Faddeev, M. Jimbo  and G. P. Pronko
for useful discussions.
We would like to thank M. Kashiwara and T. Miwa for their kind hospitality
 in RIMS.
This work is supported in part by RBRF--98--01--00070,
INTAS--96--690 (Yu.~S.). V.~F. is supported by a JSPS fellowship.

\end{document}